 \documentclass[doublespacing]{elsart}

\usepackage{natbib}
\usepackage{subfigure}
\usepackage{graphicx}
\usepackage{lscape}
\usepackage{multirow}

\usepackage{amssymb}

\begin{document}

\begin{frontmatter}



\title{Spectral hardness evolution characteristics of tracking Gamma-ray Burst
pulses}



\author{Z. Y. Peng}

\address{Department of Physics, Yunnan Normal University, Kunming 650092, P. R.
China}

\ead{pzy@ynao.ac.cn}

\author{L. Ma$^{\ast}$}
\address{Department of Physics, Yunnan Normal University, Kunming 650092, P. R.
China}

\corauth[]{Corresponding author,  astromali@126.com }

\author{R. J. Lu}
\address{
Physical Science and Technology College, Guangxi University, Nanning, Guangxi 530004, P. R. China}

\author{L. M. Fang}
\address{Department of Physics, Guangdong Institute of Education,
Guangzhou 510303, P. R. China}
\author{Y. Y. Bao}
\address{Department of Physics, Yuxi Normal
College, Yuxi 653100, P. R. China}

\author{Y. Yin}
\address{Department of Physics, Yunnan Normal University, Kunming 650092, P. R.
China}

\begin{abstract}
Employing a sample presented by Kaneko et al. (2006) and Kocevski et
al. (2003), we select 42 individual tracking pulses (here we defined
tracking as the cases in which the hardness follows the same pattern
as the flux or count rate time profile) within 36 Gamma-ray Bursts
(GRBs) containing 527 time-resolved spectra and investigate the
spectral hardness, $E_{peak}$ (where $E_{peak}$ is the maximum of
the $\nu F_{\nu}$ spectrum), evolutionary characteristics. The
evolution of these pulses follow soft-to-hard-to-soft (the phase of
soft-to-hard and hard-to-soft are denoted by rise phase and decay
phase, respectively) with time. It is found that the overall
characteristics of $E_{peak}$ of our selected sample are: 1) the
$E_{peak}$ evolution in the rise phase always start on the high
state (the values of $E_{peak}$ are always higher than 50 keV); 2)
the spectra of rise phase clearly start at higher energy (the median
of $E_{peak}$ are about 300 keV), whereas the spectra of decay phase
end at much lower energy (the median of $E_{peak}$ are about 200
keV);  3) the spectra of rise phase are harder than that of the
decay phase and the duration of rise phase are much shorter than
that of decay phase as well. In other words, for a complete pulse
the initial $E_{peak}$ is higher than the final $E_{peak}$ and the
duration of initial phase (rise phase) are much shorter than the
final phase (decay phase). This results are in good agreement with
the predictions of Lu et al. (2007) and current popular view on the
production of GRBs. We argue that the spectral evolution of tracking
pulses may be relate to both  kinematic and dynamic process even if
we currently can not provide further evidences to distinguish which
one is dominant. Moreover, our statistical results give some
witnesses to constrain the current GRB model.

\end{abstract}


\end{frontmatter}


\section{Introduction}
\label{} The origin of Gamma-ray bursts (GRBs) is still unclear even though much progress has been made, especially the recent launch of Swift.
The spectra of GRB provide the most direct information about the emission process involved in these enigmatic events. Early studies of burst
spectra showed that they vary within a given events as well as from event to event. Since the observed spectra reflect the energy content and
particle distributions within the source's emitting region, spectral variations are crucial diagnostics of underlying physical processes within
a burst, and may be a discriminant between emission mechanisms as well.

Many authors have studied the spectral evolution since the discovery of GRBs. These studies mainly focused on the ``hardness'' of bursts,
measured either by the ratio of counts in different energy channels or by more physical variables, such as the peak energy $E_{peak}$, which is
the maximum of the $\nu F_{\nu}$ spectrum. The evolution has been studied over both the entire burst, giving the overall behavior, and the
individual pulse structures, enabling us to better understand the physical mechanisms of the GRB prompt emission process.

Golenetskii et al. (1983) compared two-channel data covering $\sim$ 40-700 keV with 0.5 s time resolution from five bursts observed by the KONUS
detector on Venera 11 and Venera 12 and found that burst intensities and spectral hardness were correlated, i.e. when burst intensity increased,
the spectrum hardened. Norris et al. (1986), however, found a hard-to-soft spectral evolution trend across 10 bursts observed by the Gamma Ray
Spectrometer (GRS) and the Hard X-Ray Burst Spectrometer (HXRBS) on Solar Maximum Mission using hardness ratio. Band et al. (1992) analyzed 9
bursts observed by BATSE SDs, confirming the hard-to-soft spectral evolution. Similar result are found by many authors (e.g. Bhat et al. 1994,
Band 1997, Share \& Matz, 1998, Preece et al. 1998). Kargatis et al. (1994) studied the spectral evolution of 16 GRBs detected by Franco-Soviet
SIGNE. They found that there is no single characteristic of spectral evolution: they saw hard-to-soft, soft-to-hard, luminosity-hardness
tracking, and chaotic evolution. In the Swift era, the spectral evolution is also very ubiquitous. For example, focusing on GRBs 061121, 060614,
and 060124, Butler \& Kocevski (2007) found that the spectral evolution inferred from fitting instead models used to fit GRBs demonstrates a
common evolution-a power-law hardness-intensity correlation and hard-to-soft evolution¡ªfor GRBs and the early X-ray afterglows and X-ray
flares.

Qin et al. (2006) investigated the evolution of spectral hardness ratio of counts in different energy channels and found the evolutionary curve
of the pure hardness ratio (when the background count is not included) would peak at the very beginning of the curve, and then would undergo a
drop-to-rise-to-decay phase due to the curvature effect. Lu et al. (2007) also studied the evolution of observed spectral hardness $E_{peak}$
based on the model of highly symmetric expanding fireballs, where the Doppler effect of the expanding fireball surface is the key factor
concerned, and found that the evolutionary curve of $E_{peak}$ also undergoes a drop to rise to decay evolution.

In conclusion, such hardness parameters were typically found either to follow a ``hard-to-soft'' trend, decreasing monotonically while the flux
rises and falls, or to ``track'' the flux during an individual pulse, with the spectral hardness peaking on the leading edges of pulses (Wheaton
et al. 1973; Norris et al. 1986; Golenetskii et al. 1983; Laros et al. 1985; Kargatis et al. 1994; Ford et al. 1995).

Kaneko et al. (2006, hereafter Paper I) made a systematic spectral analysis of 350 bright GRBs observed by BATSE with high temporal and spectral
resolution. Basing on their energy fluence or peak photon flux values to assure good statistics, they selected 350 from 2704 BATSE GRBs. A
thorough analysis was performed on 350 time-integrated and 8459 time-resolved burst spectra using 5 different photon models. The Kaneko sample
is the most comprehensive study of spectral properties of GRB prompt emission to date. In the meantime, we also analyse the spectra of weak
bursts with the peak flux less than 10 photons $s^{-1} cm^{-2}$ presented by Kocevski et al. (2003) using the same ways as Paper I. Employing
the two samples we investigate the evolution of $E_{peak}$ (the maximum of the $\nu F_{\nu}$ spectrum) confining individual pulses and find that
the evolution of $E_{peak}$ within a pulse follows hard-to-soft, soft-to-hard-to-soft (flux and hardness tracking) and chaotic pattern. Similar
to Crider et al. (1997) we take the pulses as tracking if the rise and decay of $E_{peak}$ coincides with those of count rate or flux to within
1 time bin and if the rise lasts at least 3 time bins. Figure. 1 illustrates the evolutions of $E_{peak}$ for the cases of hard-to-soft and
tracking pulses, where the photon flux (middle panels) interval is from 30 kev to 2 MeV.

The pattern of hard-to-soft is the most common and the tracking evolutionary case is less common in Kaneko sample. As for the spectral evolution
of hard-to-soft pulses, many studies have been made on their origin (e.g. Liang et al. 1997). Whereas none has been made to investigate the
characteristics of spectral evolution of tracking pulses. In addition, since both Qin et al. (2006) and Lu et al. (2007) found the spectral
hardness indeed evolve in time from drop to rise to decay due to curvature effect we first wonder what the spectral evolutionary characteristics
of the tracking pulse are. We would also like to know whether the tracking pulses are interpreted by curvature effect. Hence focusing on
studying the $E_{peak}$ evolution of these tracking pulses is our main purpose in this paper. We construct this paper as follows. In $\S$ 2 the
selection of Kaneko and our sample are introduced, respectively. In $\S$ 3 we describe our analysis methods. The analysis results are presented
in $\S$ 4. The conclusions and discussion are given in the last section.

\begin{figure}
 \centering
 \resizebox{2.5in}{!}{\includegraphics{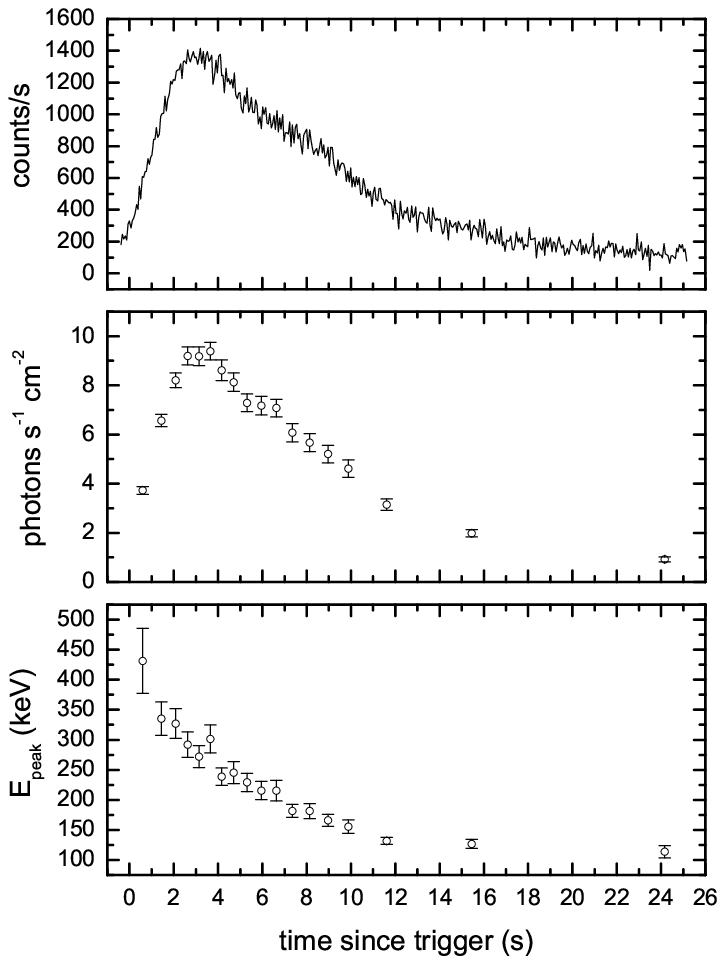}}
 \resizebox{2.6in}{!}{\includegraphics{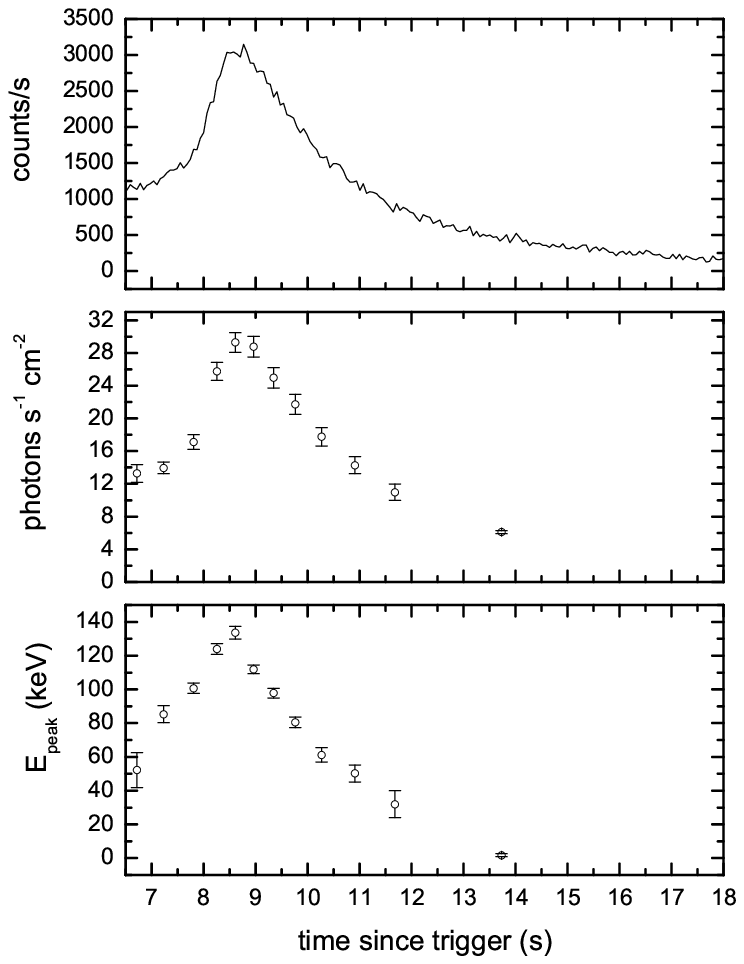}}
\caption{Example plots of evolutions of the observed peak energy
$E_{peak}$ of hard-to-soft (BATSE trigger 6397, left panel) and
tracking (BATSE trigger 2083, right panel) pulse.}
 \label{}%
 \end{figure}

\section{The Sample selection }
\subsection{The selection of Kaneko sample}
Paper I first selected 350 GRBs according to some given criterions and then made spectral analysis. In the following, we describe it simply.
(For further information about the sample, one can refer to Paper I.)
\subsubsection{The selection methodology of Kaneko sample}

a) Burst sample selection: The Kaneko sample was selected from 2704 bursts observed by BATSE. The burst selection criteria was a peak photon
flux in 256 ms (50 $\sim$ 300 keV) greater than 10 photons $s^{-1}$ $cm^{-2}$ or a total energy fluence in the summed energy range ($\sim$
20-2000 keV) larger than 2.0 $\times$ $10^{-5}$ ergs $cm^{-2}$. b) Detector selection: In order to take advantage of Large Area Detectors'
(LADs') larger effective area Paper I only selected LAD data. One of the purposes of selection one detector data was keeps the analysis more
uniform. c) Data type selection: Three LAD data types were used in Paper I. In order of priority, they were High Energy Resolution Burst data
(HERB), Medium Energy Resolution data (MER), and Continuous data (CONT). d) Time interval selection: In order to obtain the most statistical
analysis result, they used a minimum S/N of 45 for all data types. e) Energy interval selection: The lowest seven channels of HERB and two
channels of MER and CONT were usually below the electronic lower energy cutoff and were excluded. Likewise, the highest few channels of HERB and
normally the very highest channel of MER and CONT were unbounded energy overflow channels and also not usable.

\subsubsection{energy spectra analysis of Kaneko sample}
Kaneko et al. fitted 350 time-integrated spectra and 8459 time-resolved spectra adopting a set of photon models that were usually used to fit
GRB spectra.

a) Spectral fitting software

They used specific spectral analysis software RMFIT developed by BATSE team (Mallozzi, Preece \& Briggs 2005), which incorporates a fitting
algorithm MFIT that employs the forwardfolding method (Briggs 1996), and the goodness of fit is determined by $\chi^{2}$ minimization. One
advantage of MFIT is that it utilizes model variances instead of data variances, which enables more accurate fitting even for low-count data
(Ford et al. 1995).

b) Photon models

Kaneko et al. adopted 5 spectra models to fit BATSE GRB spectra. They are the Power law model, the GRB model (BAND) (Band et al. 1993), the GRB
model with fixed $\beta$ (BETA), Comptonized model (COMP) and Smoothly Broken Power Law (SBPL), respectively. Since there are many BATSE GRB
spectra that lack high-energy photons (Pendleton et al. 1997), and these no-high-energy spectra are usually fitted well with COMP model, the
only spectral model we actually use in this work is this model. The COMP model is a low-energy power law with an exponential high-energy cutoff.
It is equivalent to the BAND model without a high energy power law, the form of the COMP model is as follows:
\begin{eqnarray}
\label{eq:1}
f_{COMP}(E)=A(\frac{E}{E_{piv}})^{\alpha}\exp(-\frac{E(2+\alpha)}{E_{peak}}),
\end{eqnarray}
E$_{piv}$ was always fixed at 100 keV , therefore, the model consists of three parameters: A, $\alpha$, and $E_{peak}$.

\subsection{The selection of our sample}
First, we download the Kaneko sample from BATSE public archive (http://ww\\w.batse.msfc.nasa.gov/batse/grb/). As Paper I pointed out that the
COMP model tends to be preferable in fitting time-resolved spectra as the existence of more spectra without high-energy component (Pendleton et
al. 1997), as well as the lower S/N in each spectrum compared with the time-integrated spectra. Therefore, we can extract the time-resolved
spectra fitted with the COMP model and the following data points are excluded:

a) the resulting $\chi^{2}$ per degree of freedom is -1 when fitting each time-resolved spectrum, because in this case the nonlinear fitting of
corresponding time-resolved spectra is failure.

b) $\alpha \leq -2$ for the COMP model. As Paper I pointed out that, the fitted $E_{peak}$ represents the actual peak energy of the $\nu
F_{\nu}$ spectrum only in the case of  $\alpha \leq -2$ for the COMP model.

c) the uncertainty of its corresponding $E_{peak}$ is larger 40\% than itself. In this way, we can obtain the best statistic.

With these criterions, we can obtain useful data points of $E_{peak}$ and flux for all the bursts that Kaneko sample were selected. The count
rates data of these bursts are available in the BATSE public archive. These count rates data were gathered by BATSE's LADs, which provide
discriminator rate with 64 ms resolution from 2.048 s before the burst to several minutes after the trigger (Fishman et al. 1994).  For our
analysis, we combine the data from the four channels. Similar to Peng et al. (2006), we also subtract background from initial count rates.
Therefore, we can get the relationship between count rates and time since trigger and that flux and time since trigger as well as that between
$E_{peak}$ and the time since trigger.

Then we present the three relationships in one figure for each burst, with the upper panel showing the count rates versus time since trigger,
middle panel indicating the flux against time since trigger and the bottom one representing the $E_{peak}$ versus time since trigger. In this
manner, we can select roughly the data points of $E_{peak}$ corresponding to pulses that we are deemed ``separable'' by eyes and obtain 82
pulses. Since our work focus on the $E_{peak}$ evolution of individual pulses, we must select those pulses contaminated by other ones as few as
possible. A single functional form is used to fit these burst time profiles so that we can identified ``separable'' pulses with pulse
overlapping reduced. It is suspected that many pulses have a shapes like FRED (fast rise and exponential decay). Similar to Ryde et al. (2005)
and Peng et al. (2006), we adopt the function presented in equation (22) of Kocevski et al. (2003) (the KRL function) to fit those selected
background-subtracted pulses, combining the data from the BATSE four channels, since the function can be well-presented the FRED pulses. In
addition, a fifth parameter $t_{0}$, which measures the offset between the start of the pulse and the trigger time, is introduced. The KRL
function is
\begin{equation}
F(t)={F_{max}}(\frac{t+t_0}{t_{max}+t_0})^r[\frac{d}{d+r}+\frac{r}{d+r}(\frac{t+t_0}{t_{max}+t_0})^{(r+1)}]^{-\frac{r+d}{r+1}},
\end{equation}
where $t_{max}$ is the time of the pulse's maximum flux, $F_{max}$; r and d are the power-law rise and decay indexes, respectively. Note that
equation (2) holds for $t\geq -t_0$, when $t< -t_0$ we take $F(t)=0$.

Similar to Peng et al (2006) and Norris et al (1996), we developed and applied an interactive graphical IDL routine for fitting pulses in bursts
in order to obtain an intuitive view of the result of the fit, which allows the user to set and adjust the initial pulse parameter and the pulse
position manually before allowing the fitting routine to converge on the best-fitting model via the reduced $\chi^{2}$ minimization. The MPFIT
we used is a set of routines for robust least-squares minimization (curve fitting), using arbitrary user written IDL functions or procedures. It
is based on the well-known and tested MINPACK-1 FORTRAN package of routines available at www.netlib.org. Moreover, MPFIT functions may permit
you to fix any function parameters, as well as to set simple upper and lower parameter bounds. There are five parameters in KRL function in all.
We first set the $F_{max}$ to the 90 percent of maximum pulse intensity and the $t_{max}$ to the time of pulse's maximum intensity and then
adjust the other 3 parameter according to the pulse's shapes.

The fits are performed on the regions including a complete pulse and are examined many times to ensure that they are indeed the best ones (the
reduced $\chi^{2}$, $\chi_{\nu}^{2}$, is the minimum). In addition, the data points of $E_{peak}$ of each pulse must be larger than 6 to ensure
both of the rise and decay phase last at least 3 time bins.

In the course of fitting, we find the KRL function cannot well fit the pulses with sharp peak though it can well fit that pulses with flat peak.
Moreover, it is not proven that all pulses have the shape of FRED. Therefore, we use another function of equation (1) in Norris et al. (1996)
(the Norris function), which could be rewritten as follows:
\begin{equation}
I(t)=A\left\{
\begin{array}{ll}
(\exp(-(|t-t_{max}|/\sigma_{r})^{\nu})
&  t < t_{max},\\
(\exp(-(|t-t_{max}|/\sigma_{d})^{\nu}) & t > t_{max},
\end{array}
\right.
\end{equation}
where $t_{max}$ is the time of the pulse's maximum intensity, A; $\sigma_{r}$ and $\sigma_{d}$ are the rise (t $< t_{max}$) and decay (t $>
t_{max}$) time constant, respectively; and $\nu$ is a measure of pulse sharpness.

Norris function also have 5 parameters and combined the rise, decay time constant and pulse sharpness permit a wide variation of pulse shape. In
addition, when $\nu$ lower than unit we can yield spikier pulses.

We also first set the maximum intensity, A to the 90 percent of pulse intensity and the $t_{max}$ to the time of pulse's maximum intensity. Then
we adjust the other 3 parameter according to the pulse's shapes.

The pulses we selected are fitted with the two functions, respectively. Then we select the best fitted model parameters with smaller fitting
$\chi_{\nu}^{2}$ for each pulse, which can better present pulse profile. The pulses with fitting $\chi_{\nu}^{2}$ larger than 2 are discarded.
In this way, we obtain 34 pulses in 29 GRBs.

Since the sample presented by Kaneko et al. are bright bursts with the peak photon flux in 256 ms (50-300 keV) greater than 10 photons $s^{-1}
cm^{-2}$, we select weaker bursts with peak photon flux less than 10 photons $s^{-1} cm^{-2}$ presented by Koceviski et al. (2003) to
investigate their ${E_{peak}}$ evolutions in time because these bursts exhibit clean, single-peaked or well-separated in multipeaked events.
These burst spectral analysis is also performed by RMFIT package. We always chose the data taken with detector that are closest to line of sight
to the GRB because it has the strongest signal. We adopt the same means as Paper I to deal with these data. Due to our study focus on the
time-resolved spectra, we use, as Ryde \& Svensson (2002) did, a signal-to-noise ratio (S/N) of the observations of at least $\geq$ 30 to get
higher time resolution. We apply S/N $\sim$ 45 as much as possible since Preece et al. (1998) has shown that S/N $\sim$ 45 is needed to perform
detailed time-resolved spectroscopy. For the weak bursts, we use S/N $\sim$ 30, in which case we check that the results are consistent with
higher S/Ns. The spectra are modeled with the aforesaid COMP model. There are 34 bursts are strong enough to perform spectral analysis. Then we
also remove the data in the case of above a), b) and c). For the 34 weak bursts, only 8 pulses in 7 bursts whose ${E_{peak}}$ exhibit
soft-to-hard-to-soft spectral evolution, while the others are hard-to-soft. The trigger numbers of the 7 bursts are 1956, 3143, 4350, 5523,
5601, 6672, and 8111. We also fit them with KRL and Norris function to get best fitting parameters.

Finally we obtain a sample consisting of 42 pulses in 36 GRBs, which contains 527 time-resolved spectra in all. Presented in Table 1 are our
selected bursts, in which include BATSE trigger number, $t_{max}$, fitted $\chi_{\nu}^{2}$, FWHM (full width at half maximum), the ratio of rise
width to the decay width and the fitting function. Displayed in Figure 2 are the 4 typical examples for our selection results with two bright
bursts and 2 weak bursts, which are fitted by Norris function (the upper 2 panels) and KRL function (the bottom 2 panels), respectively. We only
afford the evolution of count rate since it can better presents time profile than flux (Ryde \& Svensson 2002) because we defined tracking as
the cases in which the hardness follows the same pattern as the flux or count rate time profile. There are four parts (for four pulses) to
Figure 2: each part is composed of two panels, with the upper panel and the bottom panel are the evolutionary curves of count rate and
$E_{peak}$, respectively. The distributions of the reduced $\chi^{2}$ for our selected sample is displayed in Figure 3.


\begin{table*}
\centering \caption{A list of burst sample with select parameters}
\begin{tabular}{cccccc }
\hline
trigger  &$\chi_{\nu}^{2}$ & T$_{max}$ & $FWHM$ & $ratio$ & fitting function\\
\hline
    676&    1.55 &    60.05  &    3.30   &  1.40   &   N\\
   1156&    1.28 &    49.29  &    18.51  &  0.85   &   K\\
   1733&    1.11 &     3.36  &    4.47   &  0.50   &   K\\
   1982&    1.38 &    15.85  &    7.02   &  0.95   &   N\\
 2083:1&    1.44 &    1.16   &    1.30   &  0.87   &   N\\
 2083:2&    1.58 &    8.68   &    2.62   &  0.52   &   K\\
 2138:1&    1.36 &    7.54   &    5.35   &  0.84   &   K\\
 2138:2&    0.99 &    78.41  &    6.06   &  0.42   &   K\\
   2156&    1.74 &    14.56  &    4.10   &  0.63   &   K\\
   2389&    1.31 &    11.67  &    23.75  &  0.57   &   N\\
   2812&   1.72  &    0.75   &    1.57   &  0.69   &   K\\
   2919&   1.43  &    0.33   &    3.26   &  0.45   &   K\\
   3003&   1.05  &    9.75   &    11.17  &  0.57   &   K\\
   3071&   1.09  &    15.90  &    15.58  &  0.82   &   N\\
   3143&   0.93  &    0.68   &    1.84   &  0.50   &   K\\
   3227&   1.56  &    101.67 &    2.72   &  0.50   &   N\\
 3415:1&   1.20  &    0.33   &    1.32   &  0.45   &   K\\


\end{tabular}
\addtocounter{table}{-1}
\end{table*}
\begin{table*}
\centering
\caption{-Continued}
\begin{tabular}{cccccc }
\hline
trigger  &$\chi_{\nu}^{2}$ & T$_{max}$ & $FWHM$ & $ratio$ & fitting function\\
\hline
   3415:2&   1.66  &    11.53  &    1.46   &  0.39   &   K\\
 3415:3&   1.60  &    44.72  &    1.09   &  1.13   &   N\\
  3491 &   1.74  &    7.74   &    1.86   &  0.42   &   N\\
  3765 &   1.40  &   66.15   &    1.65   &  0.48   &   K\\
  3891 &   1.89  &   33.26   &    0.62   &  0.31   &   K\\
  3954 &   1.11  &   0.77    &    2.87   &  0.54   &   K\\
 4350:1&   1.71  &   13.98   &    3.40   &  0.28   &   K\\
 4350:2&   1.18  &   34.11   &    6.51   &  0.52   &   K\\
   5523&   0.98  &   0.85    &    2.79   &  0.38   &   N\\
   5601&   1.31  &   7.70    &    3.74   &  0.59   &   K\\
   5621&   1.73  &   3.93    &    0.83   &  0.73   &   N\\
   5773:1&   1.33  &   8.23    &    5.923  &  0.59   &   K\\
   5773:2&   1.65  &   17.16   &    5.64   &  0.84   &   N\\
   6100  &   1.03  &   8.27    &    2.03   &  0.42   &   K\\
  6414 &   0.99  &   6.13    &    13.31  &  0.58   &   N\\
  6581 &   1.49  &  47.71    &    0.49   &  0.40   &   K\\
  6672 &   1.19  &  0.81     &    2.08   &  0.29   &   N\\
\end{tabular}
\addtocounter{table}{-1}
\end{table*}
\begin{table*}
\centering \caption{-Continued}
\begin{tabular}{cccccc }
\hline
trigger  &$\chi_{\nu}^{2}$ & T$_{max}$ & $FWHM$ & $ratio$ & fitting function\\
\hline
  6763 &   1.15  &  7.87     &    7.795  &  0.54   &   N\\
  6891 &   1.18  &  12.68    &    8.62   &  0.85   &   K\\
  7113 &   1.57  &  19.71    &    0.50   &  1.37   &   N\\
  7360 &   1.64  &  40.29    &    12.30  &  1.30   &   N\\
  7491 &   1.02  &  18.68    &    0.54   &  1.21   &   N\\
  7515 &   1.08  &  9.10     &    8.08   &  0.61   &   K\\
  7549 &   1.45  &  127.23   &    1.29   &  1.14   &   N\\
  8111 &   1.12  &  4.96     &    2.50   &  0.32   &   K\\
\hline\\

\end{tabular}

Note: N and K denotes the KRL and Norris function, respectively.

\end{table*}

\begin{figure}
\centering
\resizebox{2.5in}{!}{\includegraphics{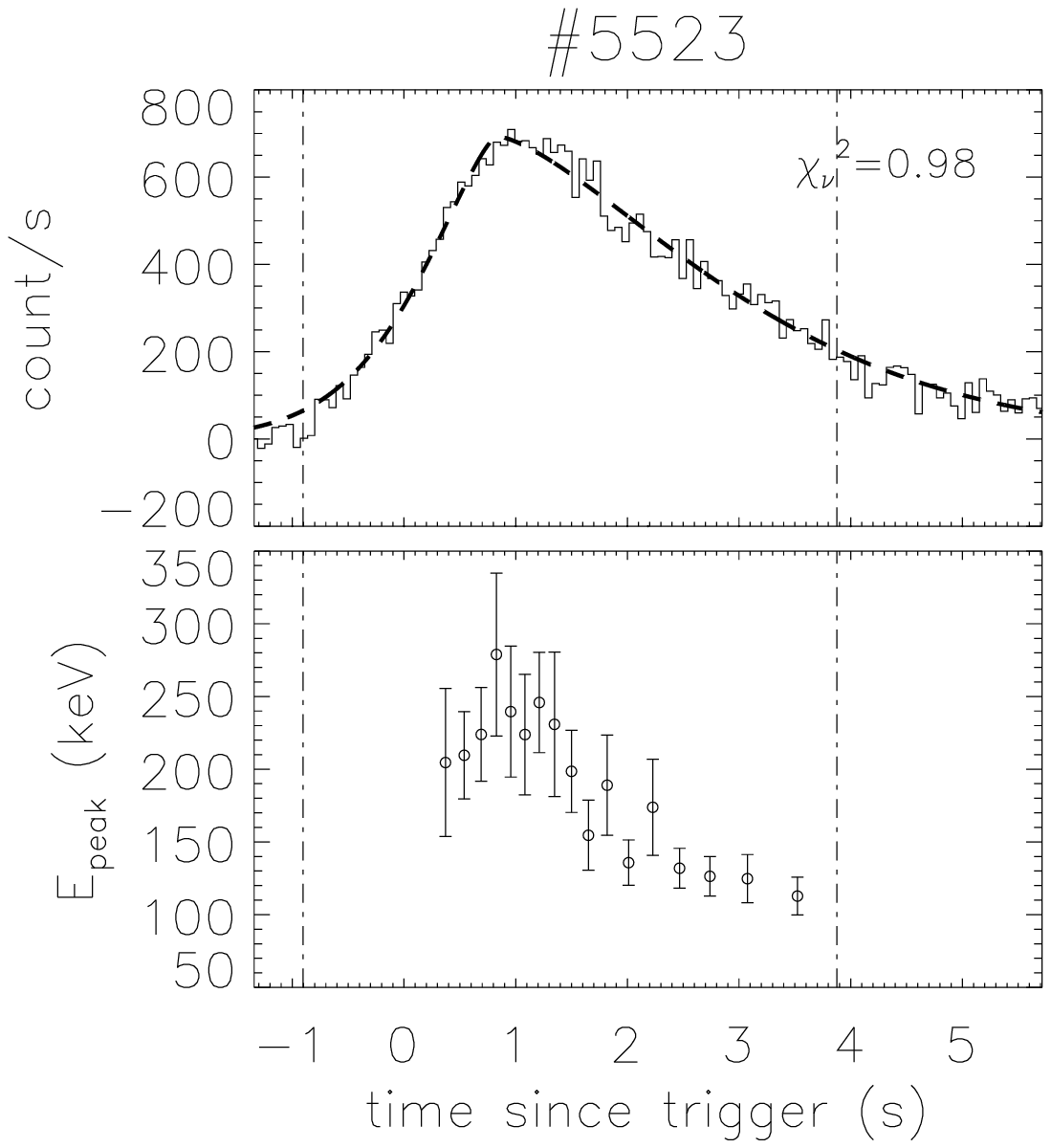}} \resizebox{2.5in}{!}{\includegraphics{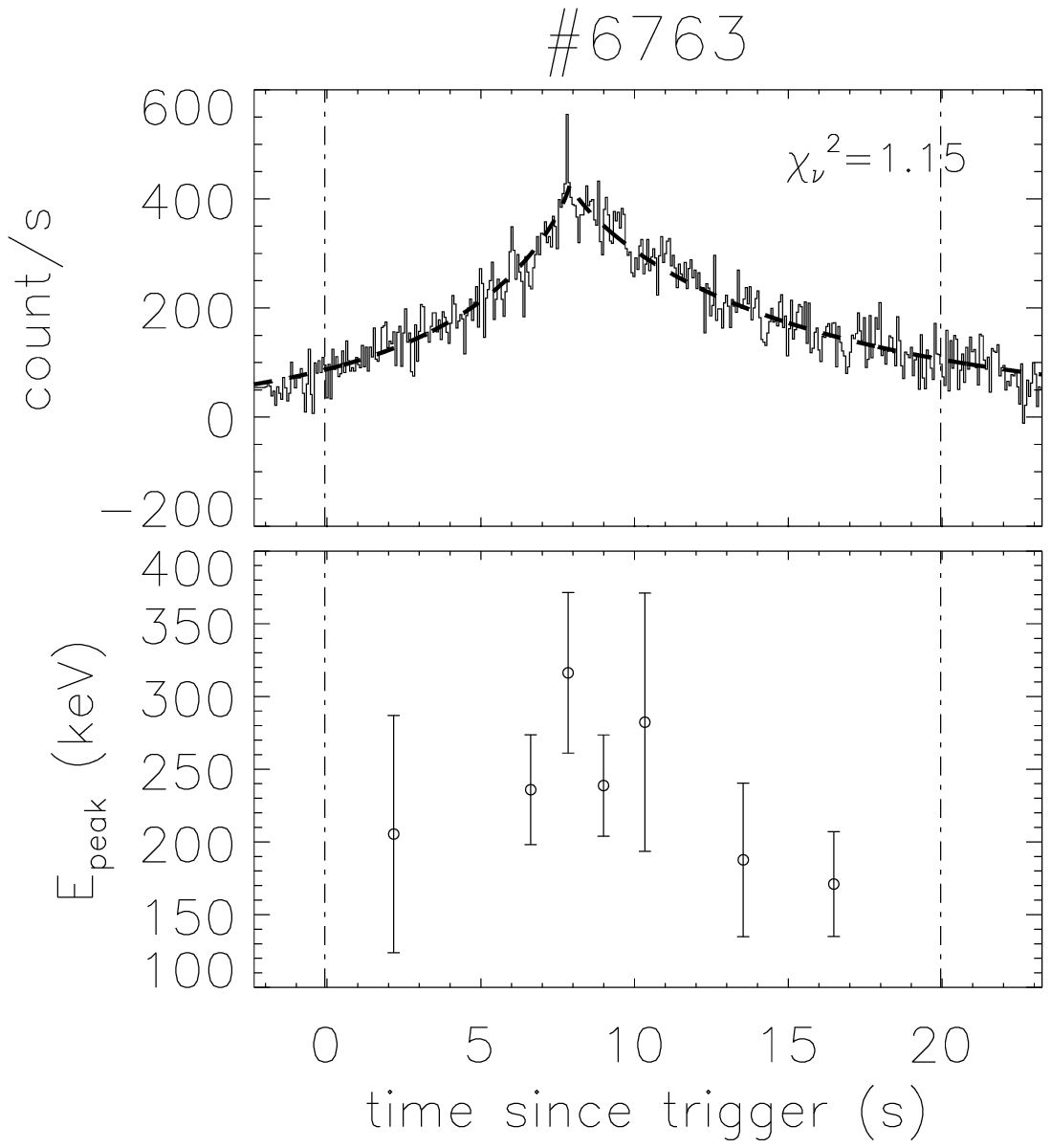}} \resizebox{2.5in}{!}{\includegraphics{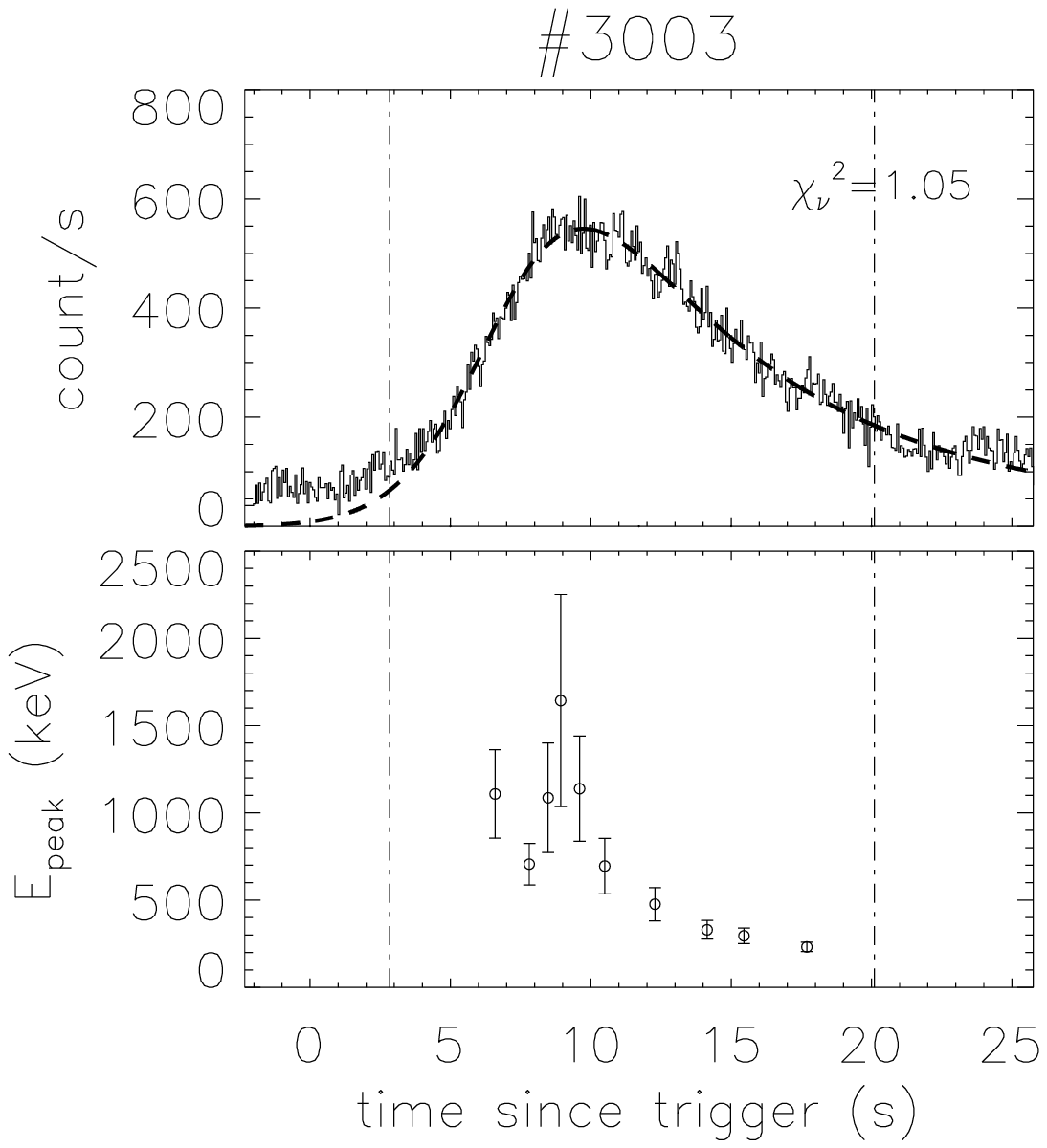}}
\resizebox{2.5in}{!}{\includegraphics{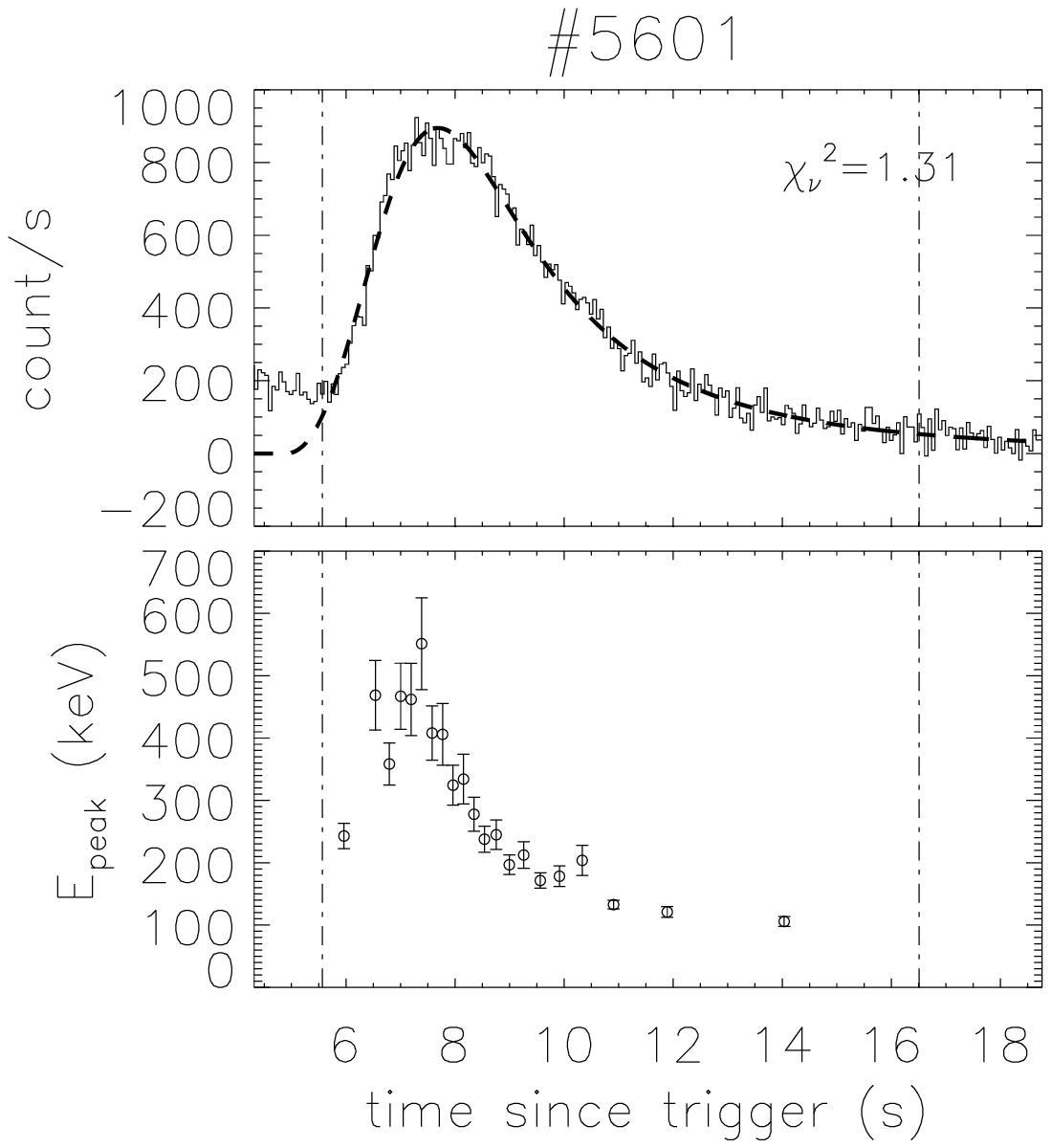}}

\caption{Example plots corresponding to BATSE trigger number 5523 (weak burst), 6763 (bright burst), 3003 (bright burst), and 5601 (weak burst)
of the evolution of count rate (top panels) and $E_{peak}$ (bottom panels) fitted by Norris function (the upper 2 panels) and KRL function (the
bottom 2 panels), where the dashed lines are the fitting curves and the dashed-dotted-dashed lines are the boundary of fitted pulses.}
\label{}%
\end{figure}


\begin{figure}
\centering \resizebox{2.5in}{!}{\includegraphics{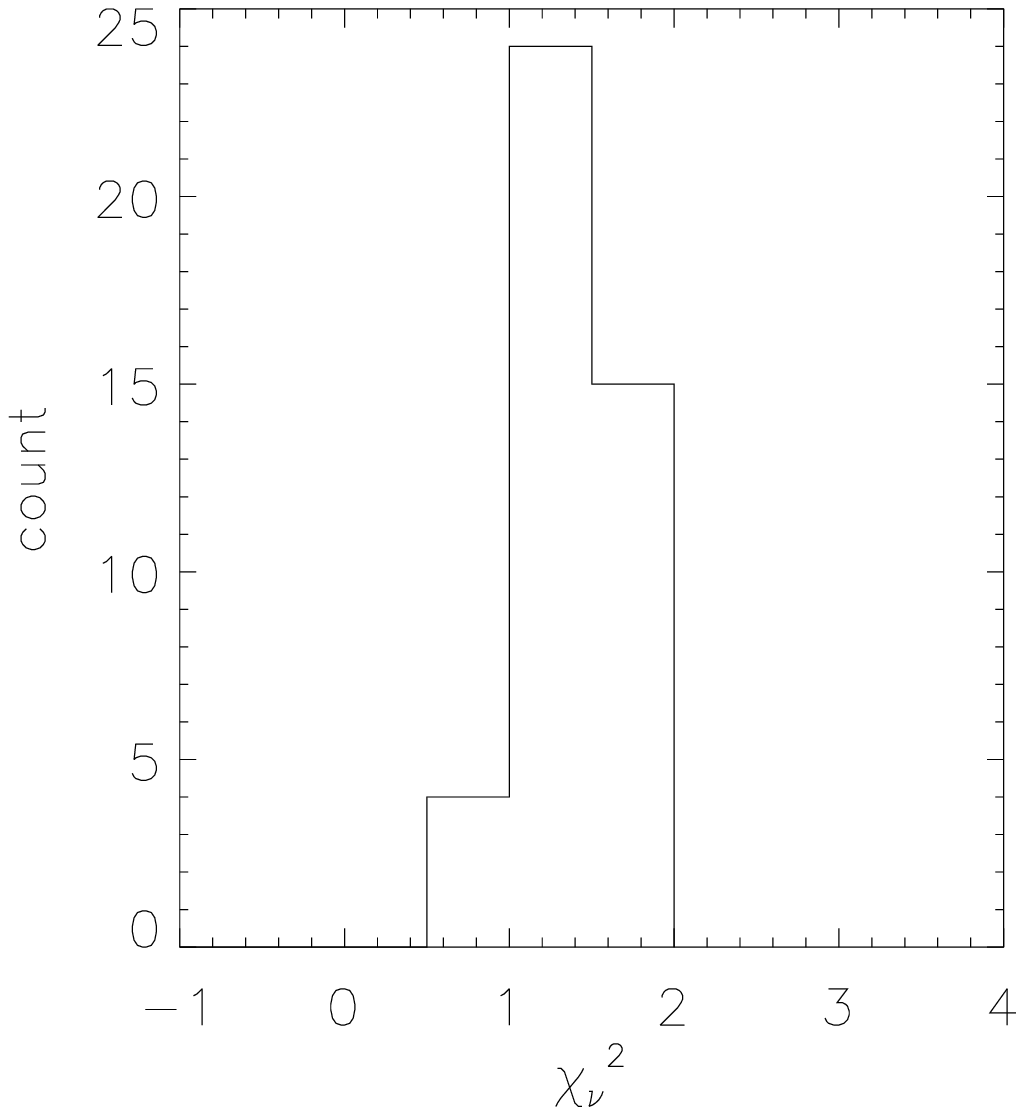}} \resizebox{2.5in}{!}{\includegraphics{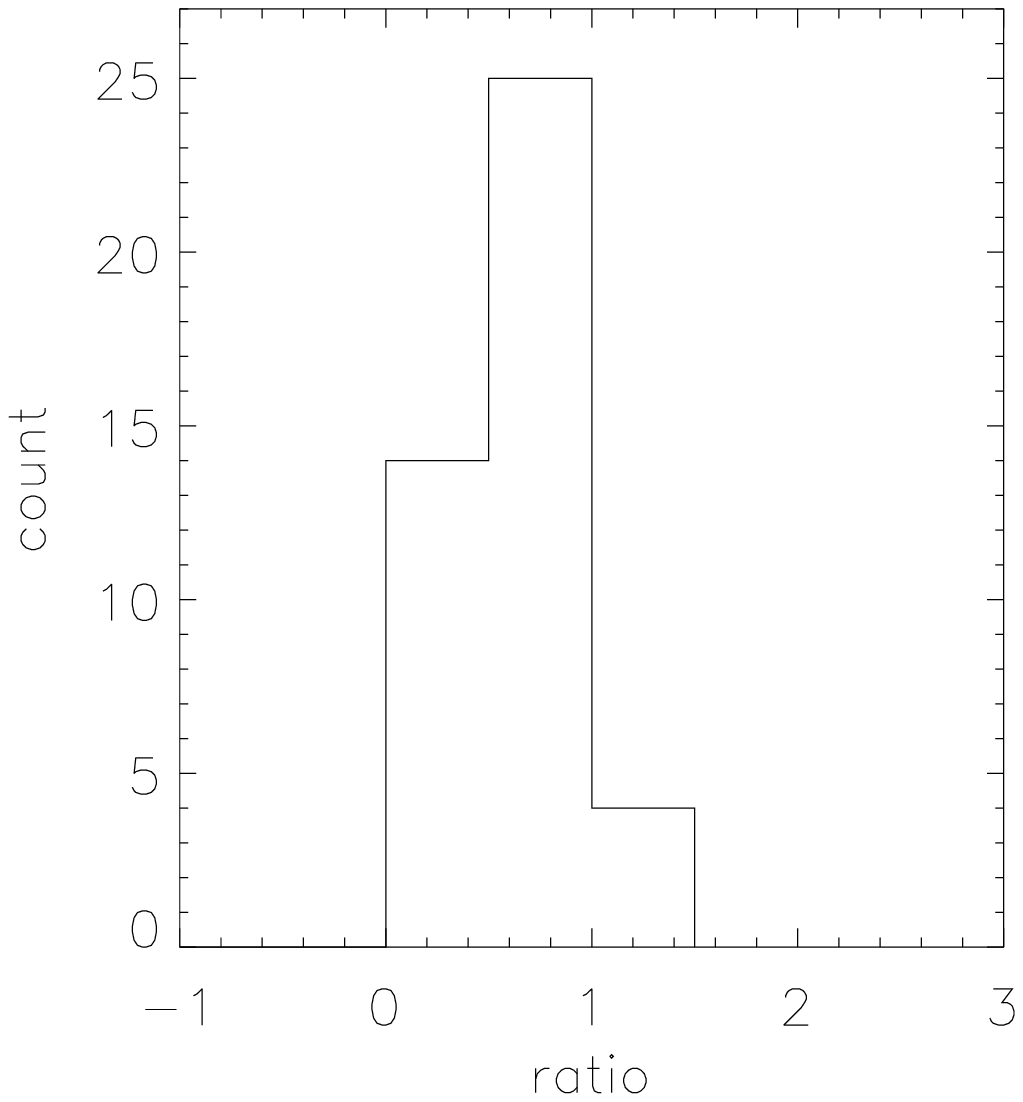}} \caption{Histograms for the
distribution of $\chi_{\nu}^{2}$ (left panel) and the ratios of the rise width to the decay width (right panel) in our selected sample.}
 \label{}%
\end{figure}

\section{analysis method}
In the previous section we have adopted KRL and Norris function to fit all background-subtracted light curves of our selected sample and then
obtained 5 fitting parameters. Therefore, the full width at half-maximum (FWHM) (see Table 1), the rise width and decay width for each pulse are
estimated. With above preparation we make the following transformation.

Firstly, For the sake of comparison, let us re-scale the time since trigger of each pulse by assigning $t_{max}$ for 0 so that the peak time of
the evolutionary curves of $E_{peak}$ almost locates at 0 and denote them as shifttime. Secondly, let us sort all these data points of
$E_{peak}$ in shifttime order and then divide them into 10 groups evenly. For the every group the histogram of $E_{peak}$ are plotted,
respectively. Thirdly, we find out the median of $E_{peak}$ of every group and indicate it by a line together with the values of 50 keV, 100
keV, 200 keV in the corresponding histogram. Fourthly, we calculate the ratios of above 200 keV, 100 keV and 50 keV, respectively, for every
group. Fifthly, we extract all the median and corresponding shifttime (here the shifttime take the middle time of start and end time for every
group). Sixthly, in order to get a uniform time we normalize shifttime in corresponding $FWHM$ of each pulse. This time are denoted as
normalizedshifttime. Then we repeat what the first to the fifth step have done. Lastly, we examine the relationship between the rise width and
the decay width of each pulse to check if these pulse profiles are different.


\section{analysis result}
\subsection{The evolutionary characteristic of $E_{peak}$ of all the pulses}

We first would like to know what the evolutionary characteristics of all the pulses are. So we study the evolution of $E_{peak}$ of 527
time-resolved spectra in shifttime and normalizedshifttime, respectively.

\begin{figure}
\centering \resizebox{2.5in}{!}{\includegraphics{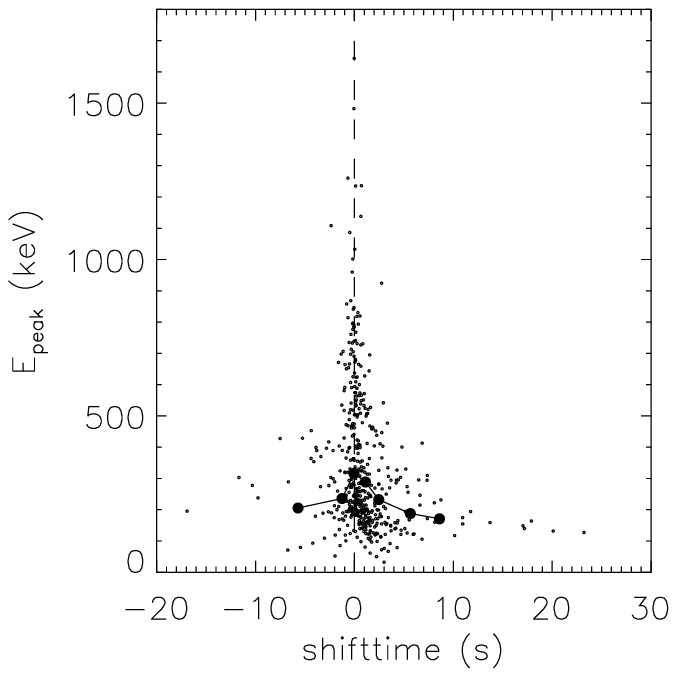}} \resizebox{2.5in}{!}{\includegraphics{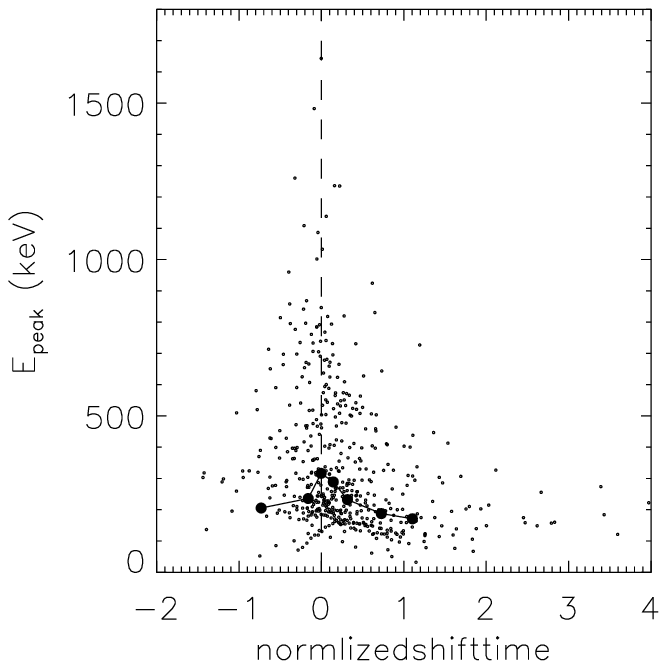}} \caption{The plots of $E_{peak}$ vs.
shifttime (left panel) and normalizedshifttime (right panel) for our selected sample, where the dot-line-dot represents the evolution of trigger
number 7515.}
 \label{}%
\end{figure}

\begin{figure}
\centering \resizebox{2.5in}{!}{\includegraphics{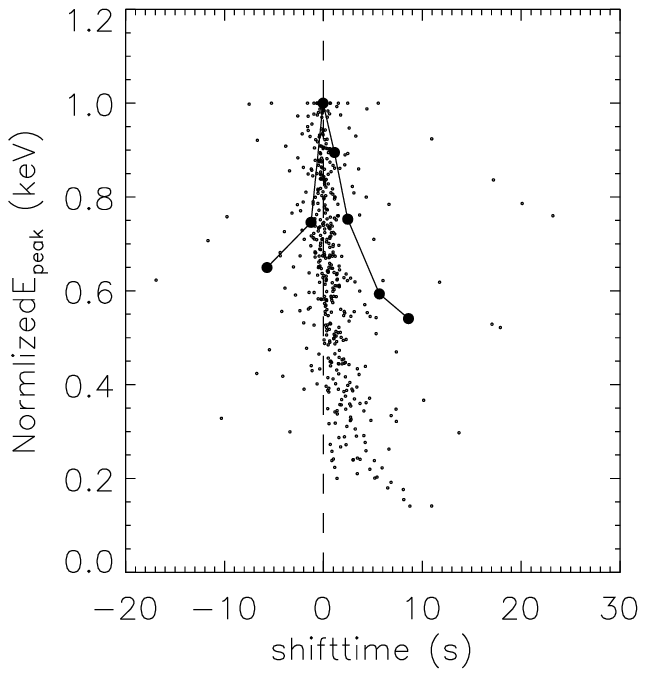}} \resizebox{2.5in}{!}{\includegraphics{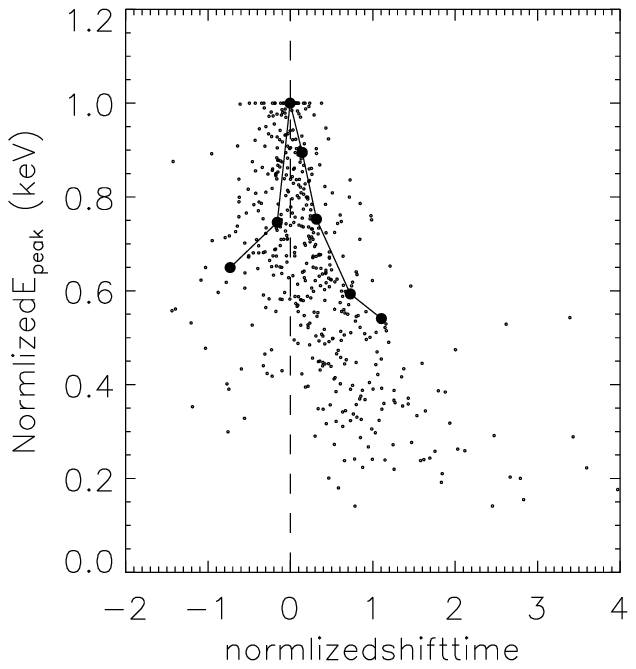}} \caption{The plots of normalized
$E_{peak}$ vs. shifttime (left panel) and normalizedshifttime (right panel) for our selected sample, where the dot-line-dot represents the
evolution of trigger number 7515.}
 \label{}%
\end{figure}
It is found in Figure 4. that the overall $E_{peak}$ evolution of our selected pulses indeed follow soft-to-hard-to-soft pattern. We also give a
example event to show how the evolution proceeds. If the case when the $E_{peak}$ are also normalized to maximum is different. We also study the
evolution of normalized $E_{peak}$ with shifttime and normalizedshifttime. The Figure 5 indicates the normalized $E_{peak}$ also follow the
pattern of soft-to-hard-to-soft. The example event clearly show the evolutionary trend.

In order to investigate the detailed characteristics of $E_{peak}$ evolution of these pulses, we divide the 527 time-resolved spectra into 10
groups evenly in the shifttime and normalizedshifttime order, respectively. Figure 6 and Figure 7 show two example (the second and the sixth
group) histograms of $E_{peak}$ to the aforesaid two sorts of time, respectively. In every panel in Figure 6 and 7 the median, 200 keV, 100 keV,
and 50 keV are indicated. In the meantime, we also give the histograms of all the $E_{peak}$ in our sample to see if their distributions are
different.

\begin{figure}
\centering \resizebox{2.5in}{!}{\includegraphics{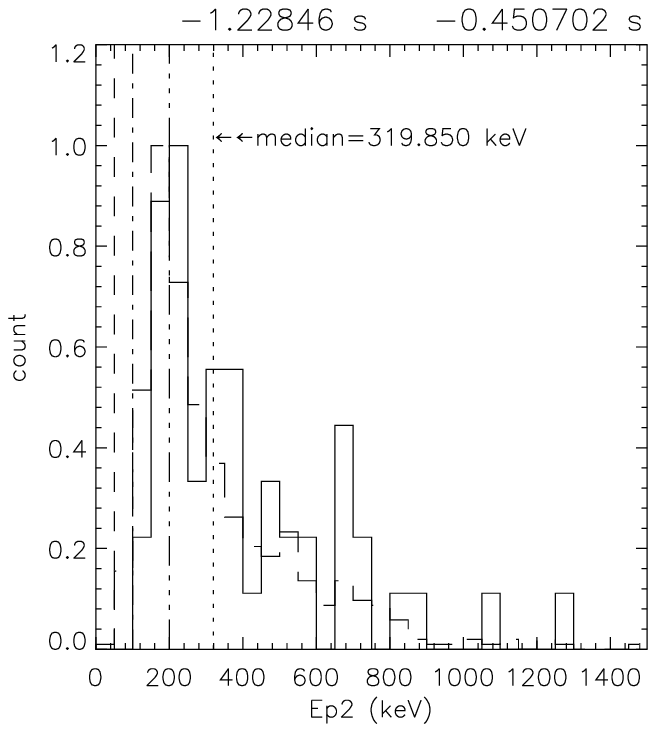}} \resizebox{2.5in}{!}{\includegraphics{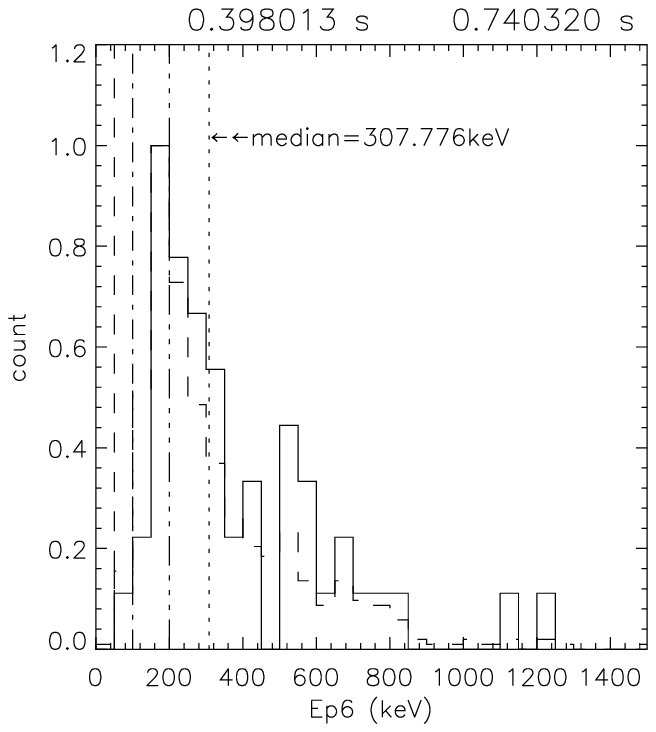}} \caption{After our sample being
divided into 10 groups in terms of shifttime, the example histograms of $E_{peak}$ (the second and the sixth group). The positions of median
(dotted lines), 200 keV (long dashed-dotted lines), 100 keV (short dashed-dotted lines) and 50 keV (dashed lines) are also plotted. where the
histograms represented by dashed lines are all the $E_{peak}$, the numbers on the top of the panels are the time interval of each group.}
 \label{}%
\end{figure}

\begin{figure}
\centering \resizebox{2.5in}{!}{\includegraphics{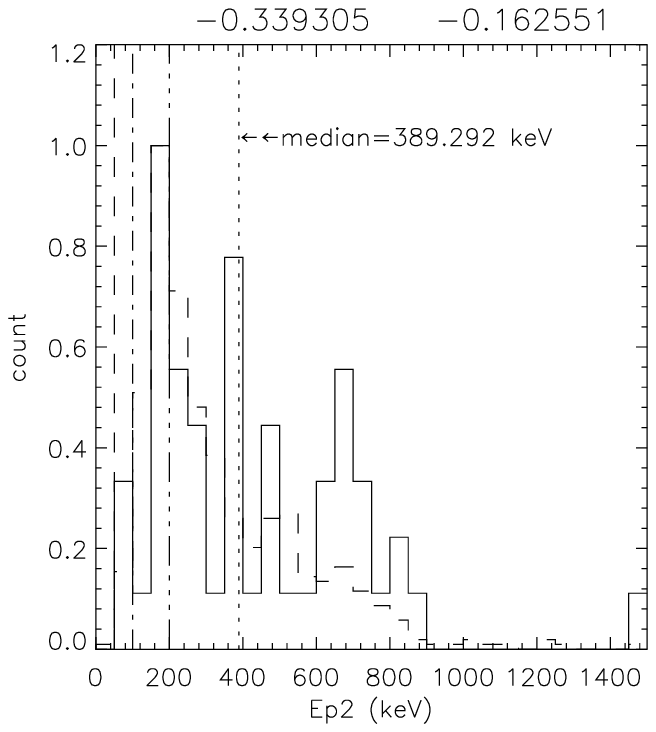}}
\resizebox{2.5in}{!}{\includegraphics{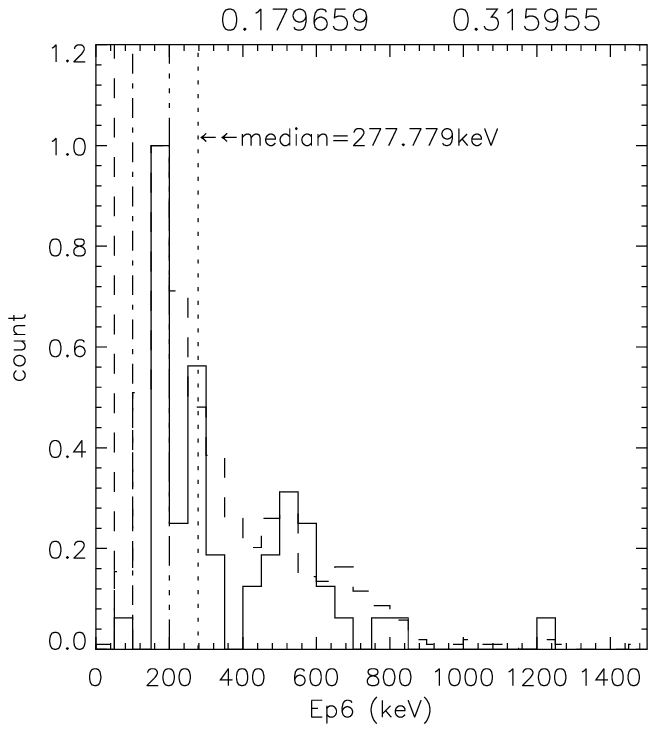}} \caption{After our
sample being divided into 10 groups in terms of normalizedshifttime,
the example histograms of $E_{peak}$ (the second and the sixth
group). The symbols are the same as those adopted in Fig. 6.}

 \label{}%
\end{figure}

Both Figure 6 and Figure 7 indicate that the evolutionary trends of
$E_{peak}$ and the changes of 200 keV, 100 keV and 50 keV. These
histograms show that the median of $E_{peak}$ first shift from low
values to high ones then to even lower than the first ones. The
variation of position of 200 keV, 100 keV, and 50 keV are also seen.

In order to obtain a more intuitive view of these points, we make the following scatter plots for all the groups that: median, the ratio of
above 200 keV, the ratio of above 100 keV and the ratio 50 keV versus shifttime, respectively. The scatter plots for the aforesaid four values
against normalizedshifttime are also made. These values are listed in Table 2 and Table 3.

\begin{figure}
 \subfigure{
 \label{fig: mini:subfig:a}
 \begin{minipage}{0.5\textwidth}
 \centering
 \includegraphics[width=3in]{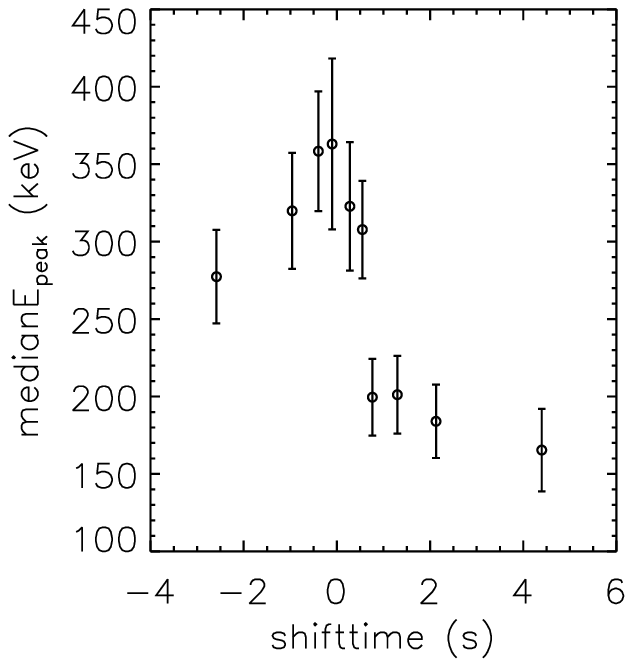}
 \end{minipage}}%
 \subfigure{
 \label{fig:mini:subfig:b}%
 \begin{minipage}{0.5\textwidth}
 \includegraphics[width=3in]{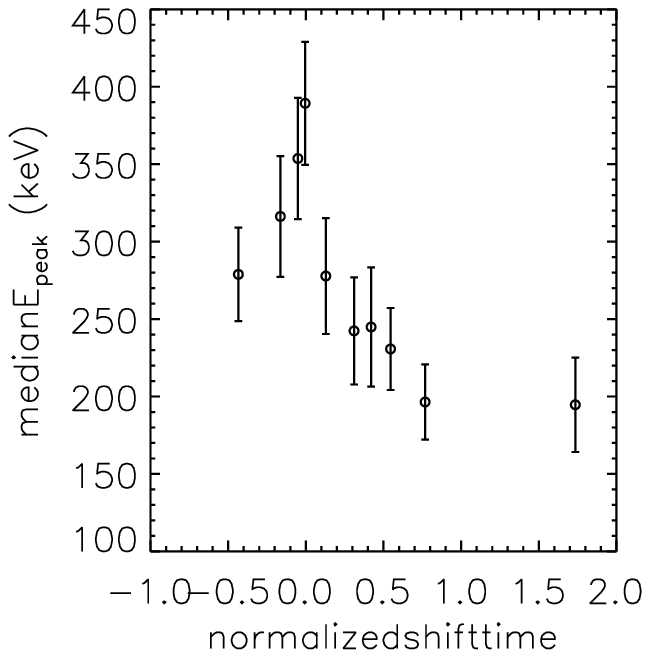}
\end{minipage}}
\caption{The plots of median of $E_{peak}$ vs. shifttime (left panel) and normalizedshifttime (right panel) after our sample having been divided
into 10 groups in terms of shifttime and normalizedshifttime.}
\label{}%
\end{figure}

\begin{figure}
 \subfigure{
 \label{fig: mini:subfig:a}
 \begin{minipage}{0.3\textwidth}
 \centering
 \includegraphics[width=2.in]{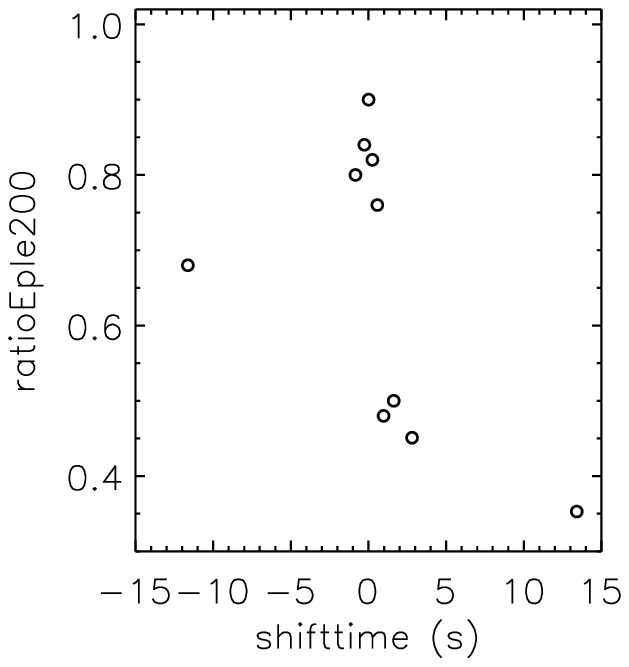}
 \end{minipage}}%
 \subfigure{
 \label{fig:mini:subfig:b}%
 \begin{minipage}{0.3\textwidth}
 \includegraphics[width=2.in]{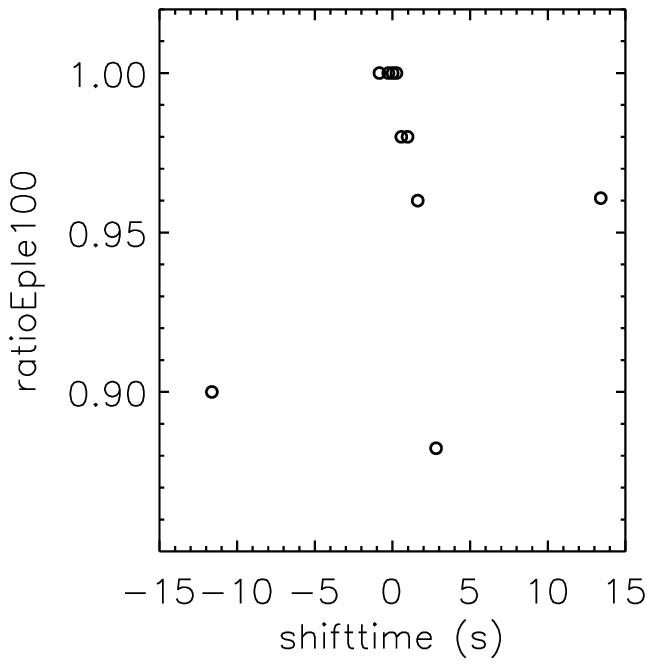}
 \end{minipage}}
 \subfigure{
 \label{fig:mini:subfig:b}%
 \begin{minipage}{0.3\textwidth}
 \includegraphics[width=2.in]{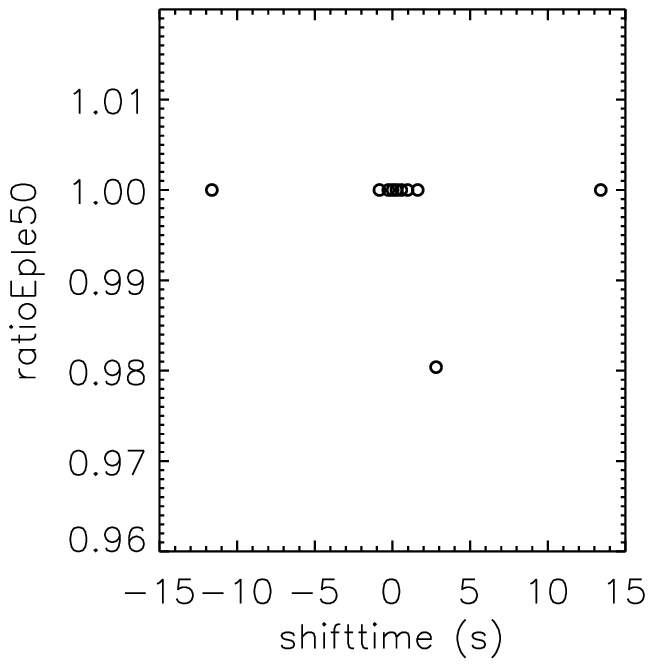}
 \end{minipage}}

 \subfigure{
 \label{fig: mini:subfig:a}
 \begin{minipage}{0.3\textwidth}
 \centering
 \includegraphics[width=2.in]{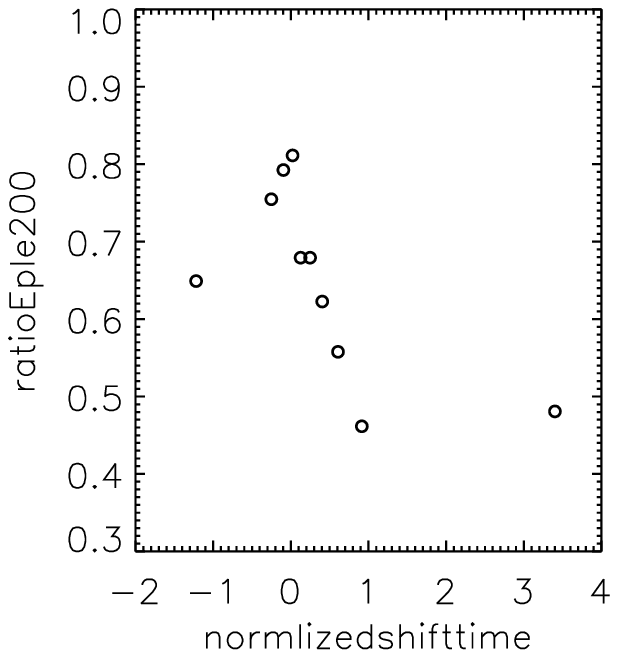}
 \end{minipage}}%
 \subfigure{
 \label{fig:mini:subfig:b}
 \begin{minipage}{0.3\textwidth}
 \includegraphics[width=2.in]{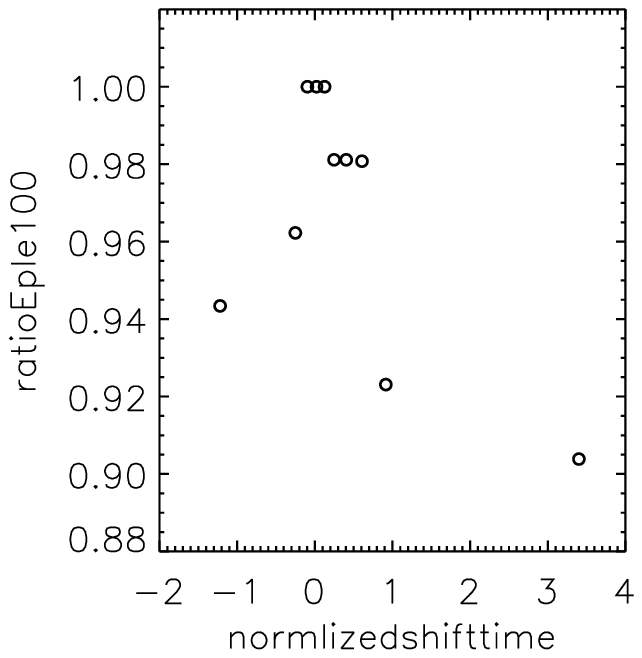}
 \end{minipage}}
 \subfigure{
 \label{fig:mini:subfig:b}%
 \begin{minipage}{0.3\textwidth}
 \includegraphics[width=2.in]{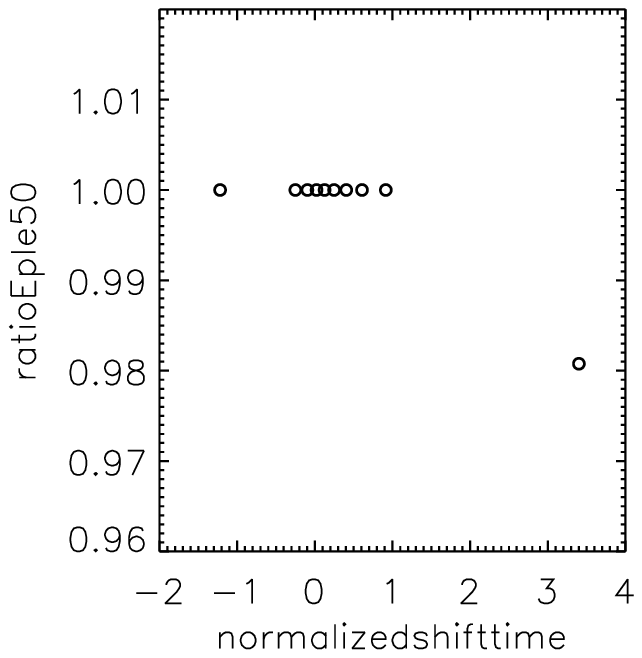}
 \end{minipage}}
\caption{The scatter plots of ratios of above 200 keV, 100 keV, and 50 keV vs. shifttime (the upper three panels) and normalizedshifttime (the
lower three panels) after being divided into 10 groups in terms of shifttime and normalizedshifttime for our selected sample. }
 \label{}%
 \end{figure}

\begin{table*}
\centering \caption{A list of the shifttime and normalizedshifttime
versus median of $E_{peak}$, respectively. }
\begin{tabular}{c|c|c|c}
\hline  shifttime (s)& medianEp (keV) & normalizedshifttime  & medianEp (keV)\\
\hline
\hline
       -2.58  &   277.36 $\pm$ 30.17 &    -0.43  &   278.24 $\pm$ 30.18 \\
\hline
       -0.96  &   319.85 $\pm$ 37.51 &    -0.16 &   316.25 $\pm$ 39.02 \\
\hline
       -0.40 &    358.37 $\pm$ 38.63 &   -0.05 &    353.46 $\pm$ 39.08\\
\hline
       -0.10 &    363.03 $\pm$ 55.17 &   -0.00 &   389.92 $\pm$  39.02\\
\hline
       0.28 &     322.78 $\pm$ 41.46 &   0.13 &    277.78 $\pm$ 37.14 \\
\hline
       0.55 &     307.78 $\pm$ 31.46 &   0.31 &   242.44 $\pm$ 34.81\\
\hline
       0.76  &    199.57 $\pm$ 34.81&   0.42  &  244.42 $\pm$ 38.47\\
\hline
       1.28 &     201.15 $\pm$ 25.13 &   0.55  &  230.96 $\pm$ 26.63\\
\hline
      2.12  &    183.98 $\pm$ 23.51  &   0.77  & 196.43 $\pm$  24.83\\
\hline
       4.39  &    165.67 $\pm$ 26.47 &   1.73  & 194.69 $\pm$  30.52\\
\hline
\end{tabular}
\end{table*}

\begin{table*}
\centering \caption{A list of the shifttime vs. the ratio of $E_{peak}$ larger than 200, 100, 50 keV and normalizedshifttime vs. the ratio of
$E_{peak}$ larger than 200, 100, 50 keV, respectively.}
\begin{tabular}{cccccccc}
\hline shifttime (s) & rEl200 & rEl100 & rEl50 & normtime &
rEl200 & rEl100 & rEl50\\
 \hline
      -11.63 &   0.68  &  0.90 &    1.00    &  -1.22  &   0.65 &    0.96   &    1.00 \\
      -0.84 &   0.80 &   1.00 &    1.00     & -0.25   &   0.75 &    0.94   &    1.00 \\
      -0.27 &  0.84 &    1.00 &    1.00     &  -0.01  &   0.79 &    1.00   &    1.00 \\
      -0.00 &   0.90  &   1.00 &   1.00     &  -0.00  &   0.81 &    1.00   &    1.00 \\
      0.25 &   0.82 &   1.00 &     1.00     &  0.13   &   0.68 &    1.00   &    1.00 \\
      0.57 &   0.76 &   0.98 &     1.00     &  0.25   &   0.68 &    0.98   &    1.00 \\
      0.97  &   0.48 &   0.96 &    1.00     & 0.40    &   0.62 &    0.98   &    1.00 \\
      1.63 &   0.50 &   0.96 &     1.00     &  0.61   &   0.56 &    0.98   &    1.00 \\
      2.81  &   0.45 &   0.88 &    0.98     &  0.91   &   0.46 &    0.92   &    1.00 \\
      13.41  &   0.35 & 0.96  &    1.00     &   3.40  &   0.48 &    0.90   &
      0.98\\
\hline\\
\end{tabular}

Note: normaltime, rEl200, rEl100 and rEl50 represent normalizedtime, ratio of $E_{peak}$ larger than 200, ratio of $E_{peak}$ larger than 100,
ratio of $E_{peak}$ larger than 50, respectively.

\end{table*}

It is clear that the evolution of median with shifttime and
normalizedshifttime are also soft-to-hard-to-soft from Figure 8,
Table 2. In addition, the phase of soft-to-hard (we denote it as
rise phase) is shorter than the phase of hard-to-soft (we denote it
as decay phase), since the time intervals of rise phase and decay
phase are 2.58 s and 4.39 s, respectively, for shifttime and 0.43,
1.73, respectively, for normalizedshifttime. The softest spectra of
rise phase (277.36 keV for shifttime and 278.24 keV for
normalizedshifttime) are harder than that of the decay phase (165.67
keV for shifttime and 194.69 keV for normalizedshifttime).

From Figure 9 and Table 3, we find that: a) the ratios of above 50
keV almost stay fixed  in the whole phase; b) the ratios of above
100 keV change from small to big at first and then to small in the
end, besides there are three bin time in the middle of the phase
remain constant. The above results show that the spectra of rise
phase are harder than 50 keV, but some spectra are softer than 100
keV. Whereas for the decay phase the softest spectra are lower than
50 keV and there are many spectra are softer than 100 keV. The
ratios of above 200 keV are, however, similar to the variation of
median, i.e. the value of $E_{peak}$ larger than 200 keV have an
asymmetrical distribution that they vary from small-to-big-to-small,
arriving at the smallest value in the end phase. Clearly, our
results are consistent with that of Preece et al. (2000), who found
that the $E_{peak}$ cluster on about 250 keV.

We do the Kolmogorov-Smirnov (K-S) test (Press et al. 1992) to check
if this $E_{peak}$ evolution is real because we would expect that
the K-S tests would also yield significant evidence that the divided
samples are different. The K-S test determines the parameter
$D_{KS}$, which measures the maximum difference in the cumulative
probability distributions over parameter space, and the significance
probability $P_{KS}$ for the value of $D_{KS}$. A small $P_{KS}$
indicates that the data sets are likely to be different (Press et
al. 1992). We employ the K-S tests between group 4 (where $E_{peak}$
is maximal) and all the groups in order to show if significant
evolution is present. Figure 10 indicates the variations of $P_{KS}$
for shifttime and normalizedshittime. It is shown in the Figure 10
that the evolution of $E_{peak}$ indeed exists.

\begin{figure}
\centering \resizebox{2.5in}{!}{\includegraphics{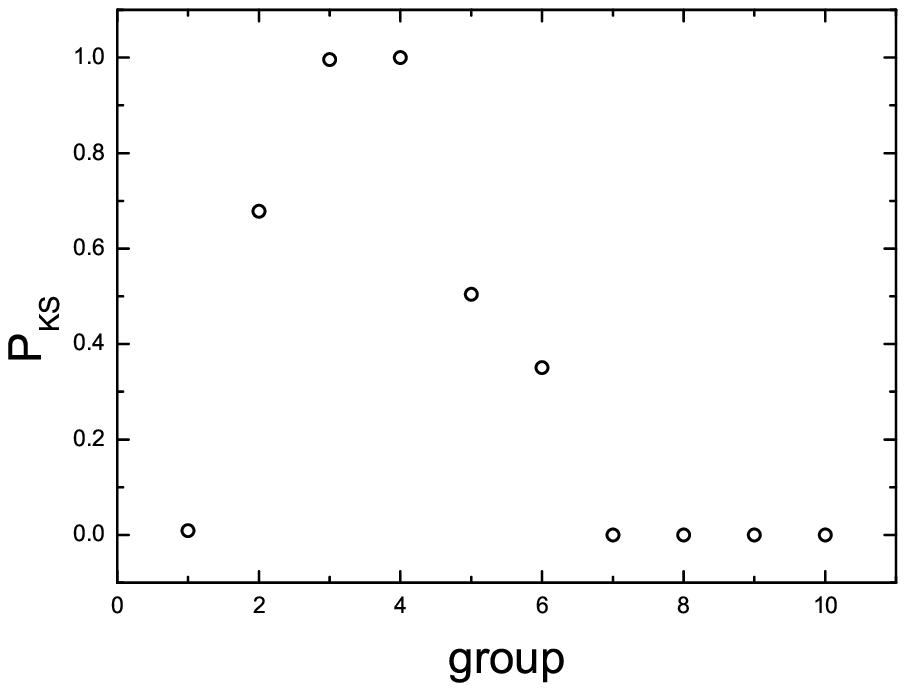}}
\resizebox{2.5in}{!}{\includegraphics{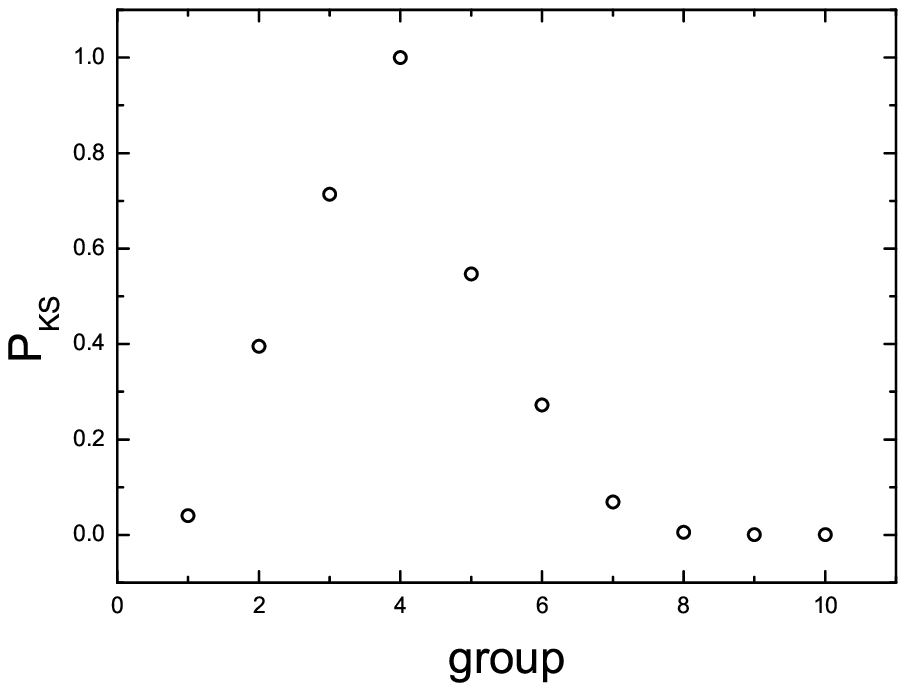}} \caption{The plots
of $P_{KS}$ and groups for the cases of shifttime (left panel) and
normalizedshifttime (right panel), where the $P_{KS}$ are the
significance probability between group 4 and all the groups after
our sample being divided into 10 groups in terms of shifttime and
normalizedshifttime.} \label{}
\end{figure}

Since many pulses have a shapes like FRED but one can not prove that
all pulses have such a shape. We check the ratios of rise width to
decay width of our selected sample (see Table 1). The ratios are
obtained by using the best model parameters due to it can reflect
the profile of corresponding pulse. As Koceviski et al. (2003)
described that KRL function is an analytical function based on
physical first principles and well-established empirical
descriptions of GRB spectral evolution. These analytical profiles
are independent of the emission mechanism and can be fully model the
FRED light curve. While Norris et al. (1996) pointed out that the
Norris function are more flexible model. It can model that pulses
with various shapes, especially for the sharp peak pulse, and would
not be constrained in the shapes of FRED. Therefore, we think that
the two models can well present the pulse shapes and can get
satisfied results. The histogram of ratios is displayed in Figure 3.
It is found that the ratios clustered at less than unit, which is
consistent with the remarks given by Norris et al. (1996) and Lee et
al. (2000a, 2000b). With the ratios less than unit we deem these
pulses are similar to FRED pulses. Since the pulse could be not
presented by only one functional form, we consider one of possible
reasons of the ratios more than unit come from the functional form,
or could be cause by the pulses overlap. It is suspected that, to
some extent, the evolution of $E_{peak}$ of tracking pulses is
related to the time profile.

\section{conclusions and discussion}

In this paper, we investigate a sample including 42 tracking pulses
within 36 GRBs involved 527 time-resolved spectra and study the
evolutionary characteristics of $E_{peak}$. The sample consist of 29
bright and 7 weak BATSE GRBs.  In order to get good statistics, we
use a S/N of the observations of at least $\sim$ 30 and arrive at
$\sim$ 45 as much as possible for the time-resolved spectroscopy.
Since the work focus on separate tracking pulse, we adopt two pulse
models to obtain better identification of the selected pulses and
discard that with large fitting $\chi_{\nu}^{2}$ ($>$2). Therefore,
we think that our sample is very representative of tracking pulses.

In order to make the time a relative uniform standard, we first make
a transformation of the time since trigger relative to the time of
maximum intensity of pulses (shifttime) and normalize the shifttime
in the width of pulse (normalizedshifttime). We find that the
evolution of $E_{peak}$ indeed follow soft-to-hard-to-soft with both
of shifttime and normalizedshifttime (see Figure 3). Then we divide
evenly our sample into 10 groups according to the two time order to
study the evolution of median as well as the ratios of above 200
keV, 100 keV and 50 keV. For this type of tracking pulse the
$E_{peak}$ of rise phase always larger than 50 keV, while some
spectra in the decay phase less than 50 keV. The spectra of rise
phase are harder than that of decay phase. In addition, we find that
the rise phase of $E_{peak}$ evolution are shorter than that of the
decay phase and this trends are established in our selected pulses.

As the previous section pointed out the $E_{peak}$ of time resolved
spectra are fitted by COMP model. Is there a bias introduced in
always using the COMP model when the Band model is the appropriate
model? Therefore we also investigate the time resolved spectra of
some pulses using Band and COMP model and then compare the values of
$E_{peak}$ when the fitting $\chi_{\nu}^{2}$ of Band model are
smaller than or comparable to that of COMP model. We find that this
lead to a little high $E_{peak}$ estimates and slightly high
$E_{peak}$'s during the rise phases.

The observed gamma-ray pulses are believed to be produced in a
relativistically  expanding and collimated fireball because of the
large energies and the short time-scales involved. To account for
the observed spectra of bursts, the Doppler effect over the whole
fireball surface (or the curvature effect) would play an important
role (e.g. Meszaros and Rees 1998; Hailey et al. 1999; Qin 2002,
2003). The Doppler model is the model describing the kinetic effect
of the expanding fireball surface on the radiation observed, where
the variance of the Doppler factor and the time delay caused by
different emission areas on the fireball surface (or the spherical
surface of uniform jets) are the key factors to be considered (for a
detailed description, see Qin 2002 and Qin et al. 2004).

Qin et al. (2006) investigated the GRBs pulses and found that the
curvature  effect influences the evolutionary curve of the
corresponding hardness ratio. They found the evolutionary curve of
the pure hardness ratio would peak at the very beginning of the
curve, and then would undergo a drop-to-rise-to-decay phase due to
the curvature effect.

Based on the model of highly symmetric expanding fireballs, Lu et
al. (2007)  investigated in detail the evolution of spectral
hardness $E_{peak}$ of FRED pulse caused by curvature effect. They
first investigated the cases that the local pulses are exponential
rise and exponential decay and exponential rise, respectively, and
found that for both of the two local pulses the evolutionary curves
of $E_{peak}$ underwent drop-to-rise-to-decay evolution, which
corresponded to A, B, and C phases, respectively. Then they assumed
that the local pulses was exponential rise and exponential decay
pulse as well as the rest frame spectra varied with time, the same
result were obtained. The B and C phase correspond to the rise phase
(soft-to-hard) and decay phase (hard-to-soft). We can also find from
Figure 1 and 3 in Lu et al. (2007) that the time interval of B phase
is shorter than that of C phase and the spectra of the B phase are
harder than that of the C phase. This situation are in good
agreement with the conclusions of the our selected pulses. Why the A
phase in our sample are not observed by BATSE? The main cause we
consider that it corresponds to the very onset of the light-curve
pulse, where the real emissions are always contaminated by the
background.

Therefore, we think the evolution of the $E_{peak}$ in our selected
pulses can be mainly caused by Doppler effect and argue that
kinematics effect may be play important role in the course of
spectral evolution.

In view of dynamics, the current popular views on the production of
GRB is  the synchrotron shock model. Based on this model, a
soft-to-hard co-moving spectrum might be come into being in the case
of a synchrotron radiation when electrons radiate at the beginning
of a shock gaining accelerations and then arriving at the maximum
speeds. The phase may be very short. After the hardest spectrum
appears, the electrons start to decelerate and the energy of
electrons become small. Moreover, the curvature effect must be at
work because the radiation come from different latitudes of fireball
(or angles of line of sight). Both of aforementioned two factors
cause the observed spectra evolute from hard to soft. The phase must
be much longer than that of the soft-to-hard (see Figure 8 and Table
2).

Kobayashi et al. (1997) discussed the possibility that GRBs result
from  internal shocks in ultrarelativistic matter and provide the
pulse profile of internal shock in Figure 1. From Figure 1 given by
Kobayshi et al. (1997) we can find that the radiation power of
internal shock indeed follow weak-to-strong-to-weak, moreover, the
time of weak-to-strong are shorter than that of the strong-to-weak.
This characteristic are consistent with that of $E_{peak}$ of the
tracking pulses, which indicate that this type of tracking pulses
also are related to the process of internal shock. Therefore, our
results for the tracking pulses do clearly support the models of GRB
shocks.

Consequently, we argue that the spectral evolution of tracking
pulses  may be relate to both of kinematic and dynamic process.
Maybe the two processes play important roles together or only one is
dominant, which are unclear and deserve the further investigation.
Our detailed statistical results of $E_{peak}$ evolution of tracking
pulses must be constrain the current theoretic model for the fact
that spectral properties of bursts can provide powerful constrains
on the detailed physical models.

In this work, we only concentrate our attention on the tracking
pulses and have not attempted to study non-tracking pulses or to
show that all pulses must be tracking pulses. We also consider and
investigate whether all pulses (or just most pulses) are
hard-to-soft or tracking. For the 34 weak burst pulses provided by
Kocevski et al. (2003) we find they are either hard-to-soft or
tracking. In addition, the fraction of tracking pulses is about 24
percent. However, we can not afford correctly the evolutionary forms
and the fraction of tracking pulses of most bright bursts because
there are many short pulses and the data points of $E_peak$ are few.
We only give the statistical properties of tracking pulses, which
will help us rule on the nature of GRB pulses as tracking.


We thank the anonymous referee for constructive suggestions and
Yi-Ping Qin  for his helpful discussions. Thanks are also given to
Rorbet Preece and Yuki Kaneko for their help with RMFIT. This work
was supported by the Natural Science Fund for Young Scholars of
Yunnan Normal University (2008Z016 ), the National Natural Science
Foundation of China (No. 10778726, 10747001), and the Natural
Science Fund of Yunnan Province (2006A0027M).





\begin{thebibliography}{}




\bibitem[Band et al., (1992)]{Ba92} Band, D., et al. 1992, AIPC, 265, 169

\bibitem[Band et al. (1993)]{Ba93}  Band, D., et al. 1993, ApJ, 413, 281

\bibitem[Band et al. (1997)]{Ba97}  Band, D., 1997, ApJ, 486, 928


\bibitem[357Bhat, 1994]{Bhat}Bhat, P. N., et al., 1994, ApJ, 426, 604

\bibitem[Briggs(1996)]{Br96}Briggs, M. S. 1996, in AIP Conf. Proc. 384, Gamma-Ray Bursts, 3rd Huntsville
Symp., ed. C. Kouveliotou, M. Briggs, \& G. Fishman (New York: AIP),
133
%
\bibitem[Butler et al. (1997)]{Bulter97}Butler, N. R., Kocevski, D., 2007, ApJ, 663, 407

%

\bibitem[Crider et al.(1997)]{Crider97}  Crider, A., et al. 1997, ApJ, 479, L39

%
%
%
%


\bibitem[Fishman et al. (1994)]{Fi94} Fishman, G., et al. 1994, ApJS, 92, 229

\bibitem[141Ford, 1995]{Ford95}Ford, L. A., et al.
1995, ApJ, 439, 307

\bibitem[353Golenetskii, 1983]{Go83}Golenetskii, S. V., et al. 1983, Nature, 306, 451

%
%
%
%
%
\bibitem[Hailey (1999)]{Ha99} Hailey, C. J., et al. 1999, ApJ, 520, L25

\bibitem[358Kaneko, 2006]{Kaneko06}Kaneko, Y., et al. 2006, ApJS, 166, 298
(Paper I)

\bibitem[356Kargatis, 1994]{Kargatis94} Kargatis, V. E., et al. 1994, ApJ, 422, 260


\bibitem[310Kobayashi, 1997]{Kobayashi97}Kobayashi, S., et al. 1997, ApJ, 490, 92

\bibitem[Kocevski et al. (2003)]{Ko03}Kocevski, D., et al. 2003, ApJ, 596, 389

\bibitem[Laros et al. (2007)]{Laros85}Laros, J. G., et al. 1985, ApJ, 290,728

\bibitem[Lee et al. (2000)]{lee00}Lee, A., et al. 2000a, ApJS, 131, 1

\bibitem[Lee et al. (2000)]{lee00}Lee, A., et al. 2000b, ApJS, 131, 21

\bibitem[Liang et al. (1997)]{Liang97}Liang, E. P., et al. 1997, ApJ, 476, L35

\bibitem[Lu et al. (2007)]{Ko07}Lu, R. J., et al. 2007, ApJ, 663, 1110


%
%



\bibitem[Nemiroff (2000)]{Nem00}Mallozzi, R. S., et al. 2005, RMFIT, A Lightcurve and
Spectral Analysis Tool, ( Huntsville: Univ. Alabama)

\bibitem[Maszaros, 1998]{Maszaros98} Maszaros, P., Rees, M. J. 1998, ApJ, 502, L105

\bibitem[355Norris, 1986]{Norris86} Norris, J. P., et al., 1986, ApJ, 301, 213

\bibitem[Norris et al. (1996)]{No96} Norris, J. P., et al. 1996, ApJ, 459, 393




\bibitem[Pendleton (1997)]{Pe97}Pendleton, G. N., et al. 1997, ApJ, 489, 175


\bibitem[Peng et al. (2000)]{Pr00}Peng, Z. Y., et al. 2006, MNRAS, 368, 1351

\bibitem[Preece et al. (1998)]{Pr98}  Preece, R. D., et al. 1998, ApJ, 496, 849

\bibitem[Preece et al. (2000)]{Pr00}Preece, R. D., et al. 2000, ApJS, 126, 19


\bibitem[Press et al. (1992)]{Pre00} Press et al. 1992, Numerical Recipes in FORTRAN (2nd ed.; New York:
Cambridge Univ. Press)

\bibitem[Qin (2002)]{Qi02}  Qin, Y.-P. 2002, A\&A, 396, 705
\bibitem[Qin (2003)]{Qi03}  Qin, Y.-P. 2003, A\&A, 407, 393
\bibitem[Qin (2004)]{Qi04} Qin, Y.-P., et al. 2004,
ApJ, 617, 439


\bibitem[Qin et al. (2006)]{Qin06} Qin, Y.-P., et al. 2006, Phys. Rev. D, 74,
063005

\bibitem[Ryde and Svensson (2002)]{Ry02}Ryde, F., Svensson, R. 2002, ApJ, 566, 210
\bibitem[Ryde and Petrosian (2002)]{Ry05}  Ryde, F., et al. 2005, A \& A,
432, 105
\bibitem[130Share, 1998]{Share98}Share, G. H., Matz, S. M., 1998, AIPC, 428, 354

\bibitem[130Share, 1998]{Wheaton73}Wheaton, W. A., et al. 1973, ApJ, 185, L57

\end{thebibliography}
\end{document}